\documentclass[onecolumn]{aastex63}
\usepackage{amsmath}
\usepackage[caption=false]{subfig}
\usepackage{threeparttable}
\usepackage{comment}
\usepackage{dcolumn}
\usepackage{bm}
\allowdisplaybreaks[1]

\submitjournal{ApJL}

\shorttitle{Cosmic Amorphous Dust}
\shortauthors{Nashimoto et al.}

\begin{document}
\title{Cosmic Amorphous Dust Model as the Origin of Anomalous Microwave Emission}

\correspondingauthor{Masashi Nashimoto}
\email{m.nashimoto@astr.tohoku.ac.jp}

\author{Masashi Nashimoto}
\affiliation{Astronomical Institute, Tohoku University, Sendai, Miyagi 980-8578, Japan}
\author{Makoto Hattori}
\affiliation{Astronomical Institute, Tohoku University, Sendai, Miyagi 980-8578, Japan}
\author{Fr{\'e}d{\'e}rick Poidevin}
\affiliation{Instituto de Astrof{\'i}sica de Canarias, E-38200 La Laguna, Tenerife, Canary Islands, Spain}
\affiliation{Departamento de Astrof{\'i}sica, Universidad de La Laguna (ULL), E-38206 La Laguna, Tenerife, Spain}
\author{Ricardo G{\'e}nova-Santos}
\affiliation{Instituto de Astrof{\'i}sica de Canarias, E-38200 La Laguna, Tenerife, Canary Islands, Spain}
\affiliation{Departamento de Astrof{\'i}sica, Universidad de La Laguna (ULL), E-38206 La Laguna, Tenerife, Spain}


\begin{abstract}
We have shown that the thermal emission of the amorphous dust composed of amorphous silicate dust (a-Si) and amorphous carbon dust (a-C) provides excellent fit both to the observed intensity and the polarization spectra of molecular clouds. 
The anomalous microwave emission (AME) originates from the resonance transition of the two-level systems (TLS) attributed to the a-C with an almost spherical shape. 
On the other hand, the observed polarized emission in submillimeter wavebands is coming from a-Si.
By taking into account a-C, the model prediction of the polarization fraction of the AME is reduced dramatically. 
Our model prediction of the 3$\sigma$ lower limits of the polarization fraction of the Perseus and W\,43 molecular clouds at 17 GHz are $8.129\times10^{-5}$ and $8.012\times10^{-6}$, respectively. 
The temperature dependence of the heat capacity of a-C shows the peculiar behavior compared with that of a-Si.
So far, the properties of a-C are unique to interstellar dust grains.
Therefore, we coin our dust model as the cosmic amorphous dust model (CAD).
\end{abstract}


\section{Introduction} \label{sec:intro}
Plenty of observations indicate that the majority of interstellar dust is composed of amorphous material ( \citealt{LiDraine2001}).
Amorphous materials show unique physical properties compared with crystalline materials.
\cite{Zeller+1971} found from laboratory measurements that the temperature dependence of the heat capacity and the thermal conductivity of amorphous materials at low temperatures shows deviation from those of crystalline materials and is linearly proportional to temperature and proportional to the square of the temperature, respectively. 
These behaviors were found universally among glasses, such as cristobalite, vitreous silica, and so on \citep{Nittke+1998}, and do not depend on the microscopic nature of the materials.
Based on these facts, \cite{Anderson+1972} and \cite{Phillips_1972} independently proposed that thermal characteristics of amorphous materials at low temperature are governed by the transition between the two-level systems (TLS) caused by the deformation of the crystal structure. 
The mechanical potential of some of the atoms composing an amorphous material becomes a double-well potential. 
Quantum mechanically, the ground state of the energy eigenstates of the atoms split into two states. 
One is described by the sum of the states trapped in each potential minimum. 
The other is described by the difference between these states. 
Small but finite energy splitting occurs between these two states. 
In the TLS model, heat absorption and heat transport are governed by the transition between these states. 
Since the TLS model successfully explained the low-temperature thermal behaviors of the amorphous materials, it has been accepted as the standard model to describe the amorphous materials. 
\cite{Paradis+2011} showed that the fact that the observed spectrum index of thermal emission from the Galactic dust from submillimeter through millimeter wavebands is smaller than 2, can be explained by taking into account the interaction between the TLS and the electromagnetic waves.

Anomalous microwave emission (AME), which shows up as an emission bump at around 10--30\,GHz, is observed ubiquitously in the various Galactic environments (see \citealt{Dickinson+2018} and references therein).
Because of the spatial correlation of the AME and the thermal dust emission, it is widely believed that the AME originates from a kind of dust \citep{Davis+2006}.
However, the physical process of its emission mechanism is still unresolved. 
Thermal emission from amorphous dust has been proposed as one of the candidates of the AME mechanism (\citealt{Jones_2009}; \citealt{Nashimoto+2020}).
Since the typical energy difference between the TLS is of the order of $\sim 1 \, \mathrm{K} \times k_\mathrm{B}$ (where $k_\mathrm{B}$ is the Boltzmann constant) \citep{Phillips_1987} that corresponds to $\sim 10 \, \mathrm {GHz} \times h$ (where $h$ is the Planck constant), the emission caused by the resonance transition between the TLS is potentially able to explain the AME. 
\cite{Nashimoto+2020} showed that the thermal emission from amorphous silicate dust (a-Si) based on the TLS model could reproduce observational features of intensity and polarization spectra from far infrared to microwave wavebands.
One of the problems of their model is that the model prediction of the polarized intensity slightly exceeds the observational upper limit of the polarized flux density obtained by QUIJOTE (\citealt{QUIJOTE1_2015}; \citeyear{QUIJOTE2_2017}). 
To date, polarized emission from the AME has not been detected.
On the other hand, it is known that silicate dust grain contributes only half of the Galactic interstellar dust, and the remaining half is composed of the carbonaceous dust grain \citep{Weingartner+2001, MishraLi2015}. 
It is worth studying whether the polarized intensity predicted by thermal emission from the amorphous dust is able to be reduced by taking into account the carbonaceous component.

In this letter, we studied whether the thermal emission from amorphous dust proposed by \cite{Nashimoto+2020} is able to provide the model consistent with the current upper limit of the polarized intensity of the AME by taking into account both a-Si and amorphous carbonaceous dust (a-C) simultaneously. 
Our model is tested by comparing observed spectral energy distributions (SEDs) from microwave through far infrared for Perseus molecular cloud and W\,43 molecular cloud. 
Structure of this letter is as follows.
In Section \ref{sec:model}, we present the amorphous dust emission model.
In Section \ref{sec:comparison}, we compare our model to the observation.
In Section \ref{sec:discuss}, we discuss the physical properties of amorphous dust predicted by the results.

\section{Model}\label{sec:model}
Thermal emission intensity and polarization spectra of amorphous dust, $I_\nu^\mathrm{d}$ and $P_\nu^\mathrm{d}$, are expressed as,
\begin{align}
    I_\nu^\mathrm{d} &=
    \sum_i N_i C_i^\mathrm{abs} B_\nu(T_i), 
    \label{eq:Inu_d} \\
    P_\nu^\mathrm{d} &=
    \sum_i N_i C_i^\mathrm{pol} B_\nu(T_i), 
    \label{eq:Pnu_d}
\end{align}
where $i$ specifies the dust species (a-Si or a-C) throughout in this paper, 
$N_i$ is the dust column density, 
$C^\mathrm{abs}_i$ is the absorption cross section, 
$C^\mathrm{pol}_i$ is the polarization cross section,
$T_i$ is the dust temperature for each species, and $B_\nu$ is the Planck function.
We assume that the shape of a dust particle is ellipsoid with $a_{x,i} \ge a_{y,i} \ge a_{z,i}$ and is characterized by the geometrical factor $L_{j,i}$ (see \citealt{Bohren+1983}):
\begin{align}
    L_{j,i} = 
    \frac{3V_i}{8\pi}
    \int^\infty_0 
    \frac{dq}
    {(q+a_{j,i}^2)\sqrt{(q+a_{x,i}^2)(q+a_{y,i}^2)(q+a_{z,i}^2)}},
    \label{eq:Lj}
\end{align}
where $j=x$, $y$, and $z$, and $V_i$ is the volume of the dust grain of species $i$. 
It is assumed that ellipsoidal dust grains of the same volume with different axial ratios are uniformly present, which is called the continuous distribution of ellipsoids (CDE), where the lower cutoff parameter $L^\mathrm{min}_i$ for $L_{x,i}$ is introduced to remove ellipsoidal dust with an extremely large axial ratio and, therefore, $L_{x,i}$ takes a value in the range of $1/3$ from  $L^\mathrm{min}_i$.
We consider the case that the minor axis of the ellipsoidal dust is perfectly aligned in a direction parallel to the interstellar magnetic field, which is assumed to be perpendicular to the line of sight. 
This assumption will be discussed in detail in Section \ref{sec:discuss}. 
The ensemble average of absorption and polarization cross sections are given as \citep{Draine+2017}:
\begin{align}
    C_i^\mathrm{abs} &=
    \frac{2\pi^2V_i}{\lambda}\mathrm{Im}\left(\chi_{0,i}^x+\chi_{0,i}^y+2\chi_{0,i}^z\right),
    \label{eq:Cabs} \\
    C_i^\mathrm{pol} &=
    \frac{2\pi^2V_i}{\lambda}\mathrm{Im}\left(\chi_{0,i}^x+\chi_{0,i}^y-2\chi_{0,i}^z\right),
    \label{eq:Cpol}
\end{align}
where $\lambda$ is the wavelength of the incident electric field, $\chi_{0,i}^x$, $\chi_{0,i}^y$, and $\chi_{0,i}^z$ are
the electric susceptibilities of the ellipsoidal dust for the incident electric field polarized along with each axis. 
These electric susceptibilities are expressed by a dielectric constant $\varepsilon_i$ for the spherical dust grains and $L_i^\mathrm{min}$ (see Equations (A15)--(A17) in \citealt{Draine+2017}; \citealt{Nashimoto+2020}).
The dielectric constant $\varepsilon_i$ is given by the electric susceptibility of the spherical dust grain, that is $\chi_{0,i}$, as
\begin{align}
    \varepsilon_i-1 = \frac{12\pi\chi_{0,i}}{3-4\pi\chi_{0,i}}.
    \label{eq:eps}
\end{align}
The electric susceptibilities are given by following equations,
\begin{align}
    \chi_{0,\mathrm{a\mathchar`-Si}} &=
    \chi_{0,\mathrm{a\mathchar`-Si}}^\mathrm{res} + 
    \chi_{0,\mathrm{a\mathchar`-Si}}^\mathrm{tun} + 
    \chi_{0,\mathrm{a\mathchar`-Si}}^\mathrm{hop} + 
    \chi_{0,\mathrm{a\mathchar`-Si}}^\mathrm{lat},
    \label{eq:chi0_sum_a-Si} \\
    \chi_{0,\mathrm{a\mathchar`-C}} &=
    \chi_{0,\mathrm{a\mathchar`-C}}^\mathrm{res} + 
    \chi_{0,\mathrm{a\mathchar`-C}}^\mathrm{tun} + 
    \chi_{0,\mathrm{a\mathchar`-C}}^\mathrm{hop} + 
    \chi_{0,\mathrm{a\mathchar`-C}}^\mathrm{lat} + 
    \chi_{0,\mathrm{a\mathchar`-C}}^\mathrm{free}.
    \label{eq:chi0_sum_a-C}
\end{align}
The first three terms in the right hand side of each equation are the TLS contributions of the electric susceptibilities where $\chi_{0,i}^\mathrm{res}$ is attributed to the resonance transition between the two levels, $\chi_{0,i}^\mathrm{tun}$ and $\chi_{0,i}^\mathrm{hop}$ describes the quantum tunneling and thermal hopping 
relaxation processes to catch up with a shift of the energy level caused by the incident of the electromagnetic waves (\citealt{Phillips_1987}; \citealt{Meny+2007}; \citealt{Nashimoto+2020}).
The contribution from the lattice vibration $\chi_{0,i}^\mathrm{lat}$ is given by the superposition of the Lorentz models.
The carbonaceous dust grain is considered to contain free electrons because a-C might be an intermediate material between conductive graphite and non-conductive diamond.
Therefore, the free electron contribution calculated by the Drude model (see, e.g., \citealt{Bohren+1983}) $\chi_{0,i}^\mathrm{free}$ is taken into account in a-C. 
In this study, the contribution from free electrons in a-C is adopted as the graphite model provided by \cite{Draine+1984}.
A basal plane could not be defined for a-C. 
Therefore, the electric susceptibilities of the perpendicular and parallel to the basal plane provided by \cite{Draine+1984} are averaged with a weight of $1:2$.

The electric susceptibilities originated from the TLS are expressed as follows \citep{Nashimoto+2020},
\begin{align}
    \chi_{0,i}^\mathrm{res} &=
    \frac{P_{0,i} |\bm{d}_{0,i}|^2}{3 \hbar}
    \int^{\Delta_{0,i}^\mathrm{max}}_{\Delta_{0,i}^\mathrm{min}} dE
    \sqrt{1 - \left(\tfrac{\Delta_{0,i}^\mathrm{min}}{E}\right)^2}
    \tanh\left(\frac{E}{2k_\mathrm{B}T_i}\right)
    \left[
        \frac{(\omega+\omega_0)\tau_+^2-i\tau_{+,i}}
        {1+(\omega+\omega_0)^2\tau_{+,i}^2}
        -\frac{(\omega-\omega_0)\tau_+^2-i\tau_{+,i}}
        {1+(\omega-\omega_0)^2\tau_{+,i}^2}
    \right],
    \label{eq:chi0_res_rev} \\
    \chi_{0,i}^\mathrm{tun} &=
    \frac{P_{0,i} |\bm{d}_{0,i}|^2}{3 k_\mathrm{B} T_i}
    \int^{\Delta_{0,i}^\mathrm{max}}_{\Delta_{0,i}^\mathrm{min}} dE
    \int^1_{\tfrac{\Delta_{0,i}^\mathrm{min}}{E}} du
    \frac{\sqrt{1-u^2}}{u}
    \mathrm{sech}^2 \left(\frac{E}{2k_\mathrm{B}T_i}\right)
    \frac{1+i\omega\tau_{\mathrm{tun},i}(E,u)}{1+\omega^2\tau_{\mathrm{tun},i}^2(E,u)},
    \label{eq:chi0_tun_rev} \\
    \chi_{0,i}^\mathrm{hop} &=
    \frac{P_{0,i} |\bm{d}_{0,i}|^2}{3 k_\mathrm{B} T_i}
    \int^{\Delta_{0,i}^\mathrm{max}}_{\Delta_{0,i}^\mathrm{min}} dE \,
    \mathrm{sech}^2 \left(\frac{E}{2k_\mathrm{B}T_i}\right)
    \left[
        \ln
        \left(\tfrac{E}{\Delta_{0,i}^\mathrm{min}}
        +\sqrt{\left(\tfrac{E}{\Delta_{0,i}^\mathrm{min}}\right)^2 - 1}      
        \right)
        -\sqrt{1 - \left(\tfrac{E}{\Delta_{0,i}^\mathrm{min}}\right)^2}
    \right]
    \nonumber \\ & \ \ \ \times
    \int^\infty_0 dV_0\, f(V_0)\,
    \frac{1+i\omega\tau_{\mathrm{hop},i}(V_0)}{1+\omega^2\tau_{\mathrm{hop},i}^2(V_0)},
    \label{eq:chi0_hop_rev} \\
    P_{0,i} &=
    \frac{n_i f_i^\mathrm{TLS}}{\Delta_{0,i}^\mathrm{max}}
    \left[ 
    \ln \left( \frac{
        \sqrt{(\Delta_{0,i}^\mathrm{max})^2-(\Delta_{0,i}^\mathrm{min})^2} + \Delta_{0,i}^\mathrm{max}}
        {\Delta_{0,i}^\mathrm{min}}
    \right)
    - \sqrt{
        1-\left(\frac{\Delta_{0,i}^\mathrm{min}}{\Delta_{0,i}^\mathrm{max}}\right)^2}
    \right],
    \label{eq:P0} 
\end{align} 
where $\Delta_{0,i}^\mathrm{max}$ and $\Delta_{0,i}^\mathrm{min}$ are the maximum and the minimum of the tunneling splitting energy $\Delta_0$, 
$\bm{d}_{0,i}$ is the expectation value of the electric dipole moment at the potential minimum for an atom,
$\tau_{+,i}$ is the dephasing time, 
$\tau_{\mathrm{tun},i}$ and $\tau_{\mathrm{hop},i}$ are the relaxation time for the tunneling and the hopping, respectively (see \citealt{Meny+2007}; \citealt{Nashimoto+2020}),
$n_i$ is the atomic number density of a dust grain, 
$f^\mathrm{TLS}_i$ is a fraction of atoms showing the TLS for each dust species,
$E=\hbar\omega_0$ is the energy splitting of the TLS, 
$u$ is the ratio of $\Delta_0$ to $E$, 
$V_0$ is the height of the potential barrier,
$f(V_0)$ is the distribution function of $V_0$ modeled by the Gaussian with the mean of 550\,K$\times k_\mathrm{B}$ and the deviation of 410\,K$\times k_\mathrm{B}$, and 
$\omega$ is the angular frequency of the incident electric field.
We assume that the dephasing time of a-Si, $\tau_{+,\mathrm{a\mathchar`-Si}}$, is much longer than $\omega_0^{-1}$. 
This is equivalent to assume that the resonance transition probability corresponding to the energy scale between the two levels is extremely small. 
Under these assumptions, $\chi_{0,\mathrm{a\mathchar`-Si}}^\mathrm{res}$ is reduced to: 
\begin{align}
    \mathrm{Re} 
    \left(\chi_{0,\mathrm{a\mathchar`-Si}}^\mathrm{res}\right) &\simeq
    \frac{P_{0,\mathrm{a\mathchar`-Si}} |\bm{d}_{0,\mathrm{a\mathchar`-Si}}|^2}{3 \hbar}
    \int^{\Delta_{0,\mathrm{a\mathchar`-Si}}^\mathrm{max}}_{\Delta_{0,\mathrm{a\mathchar`-Si}}^\mathrm{min}} dE
    \sqrt{1 - \left(\tfrac{\Delta_{0,\mathrm{a\mathchar`-Si}}^\mathrm{min}}{E}\right)^2}
    \tanh\left(\frac{E}{2k_\mathrm{B}T_\mathrm{a\mathchar`-Si}}\right)
    \frac{-2\omega}{\omega^2-\omega_0^2},
    \label{eq:chi0_res_approx_re} \\
    \mathrm{Im} 
    \left(\chi_{0,\mathrm{a\mathchar`-Si}}^\mathrm{res}\right) &\simeq
    \left\{ \begin{array}{cl}
    \frac{\pi P_{0,\mathrm{a\mathchar`-Si}} |\bm{d}_{0,\mathrm{a\mathchar`-Si}}|^2}{3}
    \sqrt{1 - \left(\tfrac{\Delta_{0,\mathrm{a\mathchar`-Si}}^\mathrm{min}}{\hbar\omega}\right)^2}
    \tanh\left(\frac{\hbar\omega}{2k_\mathrm{B}T_\mathrm{a\mathchar`-Si}}\right) &
    \ ;\ \Delta_{0,\mathrm{a\mathchar`-Si}}^\mathrm{min} \le \hbar\omega \le \Delta_{0,\mathrm{a\mathchar`-Si}}^\mathrm{max} \\
    0 & \ ;\ \mathrm{otherwise}.
    \end{array} \right.
    \label{eq:chi0_res_approx_im}
\end{align}
Equation \eqref{eq:chi0_res_approx_im} coincides with the formula presented by \cite{Meny+2007}.

Since, in our model, the main contributor in the frequency range beyond the infrared are big grains, we neglect the size distribution of the dust grains and the dust size is fixed to $a_i=0.1\,\mu$m where $a_i \equiv (a_{x,i} a_{y,i} a_{z,i})^{1/3}$.
We can safely assume that a big grain stays at the temperature defined by the equilibrium between the heating by the interstellar radiation field (ISRF) and the radiative cooling. 
To calculate the equilibrium temperature of each species, the following relations provided by \cite{Tielens_2005} are adopted:
\begin{align}
    T_\mathrm{a\mathchar`-Si} &=
    13.5 \left(\frac{0.1\,\mu\mathrm{m}}{a_\mathrm{a\mathchar`-Si}}\right)^{0.06}
    \left(\frac{G_0}{1.7}\right)^{1/6},
    \label{eq:T_a-Si} \\
    T_\mathrm{a\mathchar`-C} &=
    15.7 \left(\frac{0.1\,\mu\mathrm{m}}{a_\mathrm{a\mathchar`-C}}\right)^{0.06}
    \left(\frac{G_0}{1.7}\right)^{1/5.8},
    \label{eq:T_a-C}
\end{align}
where $G_0$ is the scale factor of the ISRF, and power law dependence of wavelength for the long-wavelength dust opacity with power law index of 2 for a-Si and 1.8 for a-C are assumed \citep{Tielens_2005}.
Although the dust opacity in our model shows complex wavelength dependence far from the power law model and the above relations are not self-consistent with our dust opacity model, the adopted relations can be used as good estimator of dust temperature as discussed in Section \ref{sec:discuss}.
Since the wavelength dependence of dust opacity in our model depends on the dust temperature, adopting the above relations dramatically reduces fitting cost. 
In this study, $G_0$ is treated as one of the fitting parameters where $G_0=1.7$ represents the average value of the ISRF in the Galactic interstellar space.
A relative abundance of a-C to a-Si in number is fixed to reproduce the accumulative dust mass ratio $M_\mathrm{a\mathchar`-Si}/M_\mathrm{a\mathchar`-C}$ of about 1.2 given by \cite{Hirashita+2009}.
The total column density of the dust defined by $N_\mathrm{d}=N_\mathrm{a\mathchar`-Si}+N_\mathrm{a\mathchar`-C}$, is treated as a fitting free parameter.
For a-Si, $f^\mathrm{TLS}_\mathrm{a\mathchar`-Si}$ and $L^\mathrm{min}_\mathrm{a\mathchar`-Si}$ are fitting free parameters. 
The $\Delta^\mathrm{max}_{0,\mathrm{a\mathchar`-Si}}$, $\Delta^\mathrm{min}_{0,\mathrm{a\mathchar`-Si}}$ and $\tau^+_\mathrm{a\mathchar`-Si}$ of a-Si are set to reproduce the model proposed by \cite{Meny+2007}.
In the case of a-C, $\Delta^{\rm max}_{0,\mathrm{a\mathchar`-C}}$, $\Delta^{\rm min}_{0,\mathrm{a\mathchar`-C}}$ and $\tau^+_\mathrm{a\mathchar`-C}$ are treated as fitting free parameters additional to $f^\mathrm{TLS}_\mathrm{a\mathchar`-C}$ and $L^\mathrm{min}_\mathrm{a\mathchar`-C}$.
The adopted values of the physical parameters for a-Si and a-C are summarized in Table \ref{tab:fixed-params}.
The tunneling relaxation time $\tau_{\mathrm{tun},i}$ is evaluated from the mass density of a dust grain $\rho_i$, the sound velocity for transverse waves $c_{\mathrm{t},i}$, and the elastic dipole for transverse waves $\gamma_{\mathrm{t},i}$ by using the formula described in \cite{Phillips_1987}; \cite{Meny+2007}; \cite{Nashimoto+2020}. 
We assume that the pre-exponential factor for hopping relaxation time $\tau_{\mathrm{hop},i}^0$ have the same value for each dust species.
\begin{table}
    \centering
    \caption{Fixed parameters' values}
    \label{tab:fixed-params}
    \begin{tabular}{lcccc}
        \hline\hline
        Parameter & 
        \multicolumn{2}{c}{Value} & 
        \multicolumn{2}{c}{Ref.$^\ast$} \\
        & a-Si & a-C & a-Si & a-C \\ 
        \hline
        $\rho_i$ ($\mathrm{g\, cm^{-3}}$) & 
        3.5 & 1.6 & 1 & 2 \\
        $n_i$ ($\mathrm{cm^{-3}}$) & 
        $8.6\times10^{22}$ & $7.8\times10^{22}$ & 1 & 2 \\
        $M_i/(\sum_i M_i)$ &
        0.54 & 0.46 & 3 & 3 \\
        $c_{\mathrm{t},i}$ ($\mathrm{cm\, s^{-1}}$) & 
        $3\times10^5$ & $2.4\times10^6$ & 4 & 2 \\ 
        $\gamma_{\mathrm{t},i}$ (eV) &
        1 & 1 & 4 & ---$^\dagger$ \\
        $|\bm{d}_{0,i}|$ (D) &
        1 & 1 & 5 & ---$^\dagger$ \\
        $\tau_{\mathrm{hop},i}^0$ (s) &
        $10^{-13}$ & $10^{-13}$ & 4 & ---$^\dagger$ \\
        $\Delta_{0,i}^\mathrm{max}/\hbar$ ($s^{-1}$) &
        $2\times10^{13}$ & free$^\ddagger$ & 5 & --- \\
        $\Delta_{0,i}^\mathrm{min}/k_\mathrm{B}$ (K) &
        $2\times10^{-3}$ & free$^\ddagger$ & 6 & --- \\
        $\tau_{+,i}$ (s) &
        $\infty$ & free$^\ddagger$ & 5 & --- \\
        \hline
    \end{tabular}
    \begin{tablenotes}\footnotesize
        \item[] $\ast$ References: 1. \cite{Li+2001}; 2. \cite{Wei+2005}; 3. \cite{Hirashita+2009}; 4. \cite{Bosch_1978}; 5. \cite{Meny+2007}; 6. \cite{Phillips_1987}.
        \item[] $\dagger$ These are applied to the same value of a-Si because there is no reference.
        \item[] $\ddagger$ These are regarded as free parameters.
    \end{tablenotes}
\end{table}

\section{Comparison with Observations}\label{sec:comparison}
\begin{figure*}[!t]
    \centering
    \subfloat[][Perseus]{
      \includegraphics{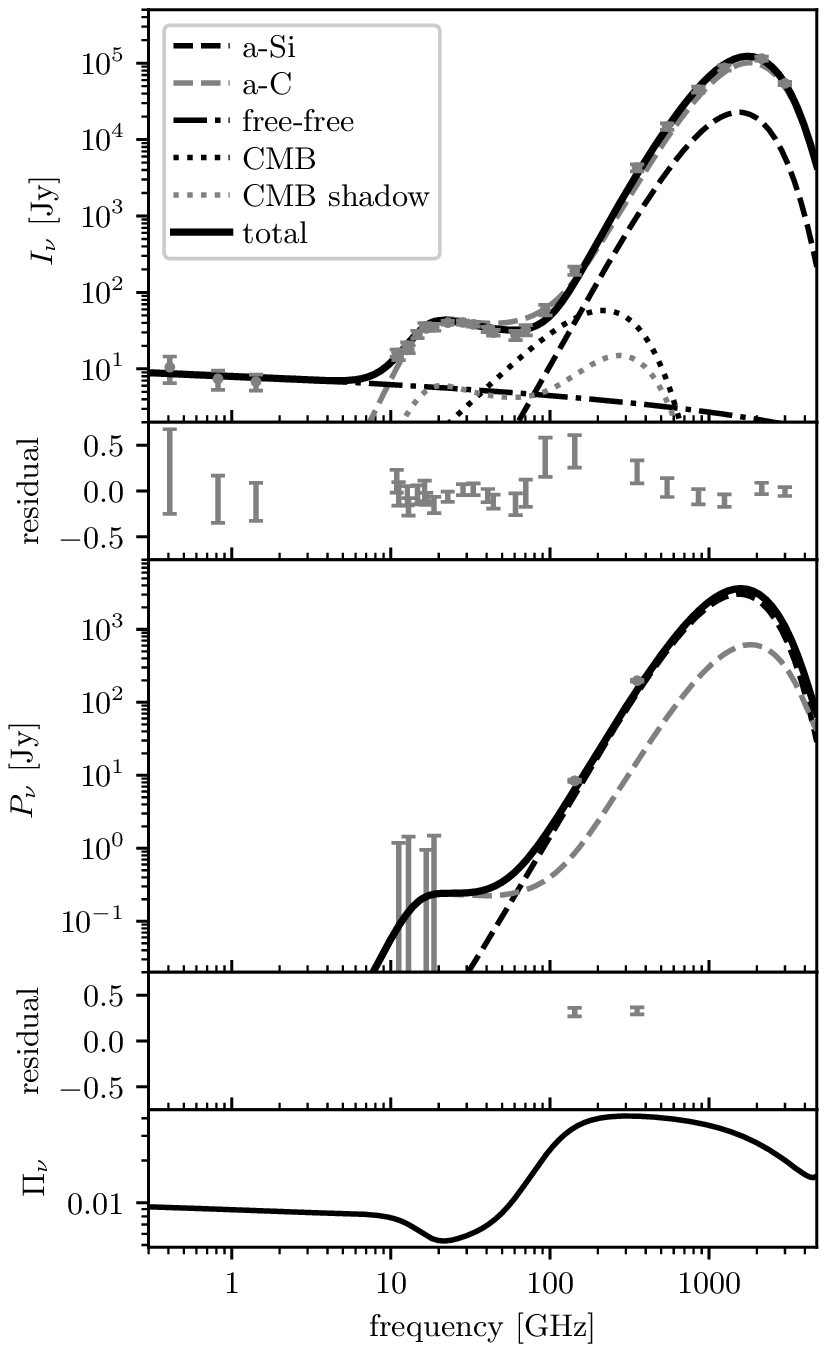}
      \label{fig:SED_per}
    }
    \quad
    \subfloat[][W\,43]{
      \includegraphics{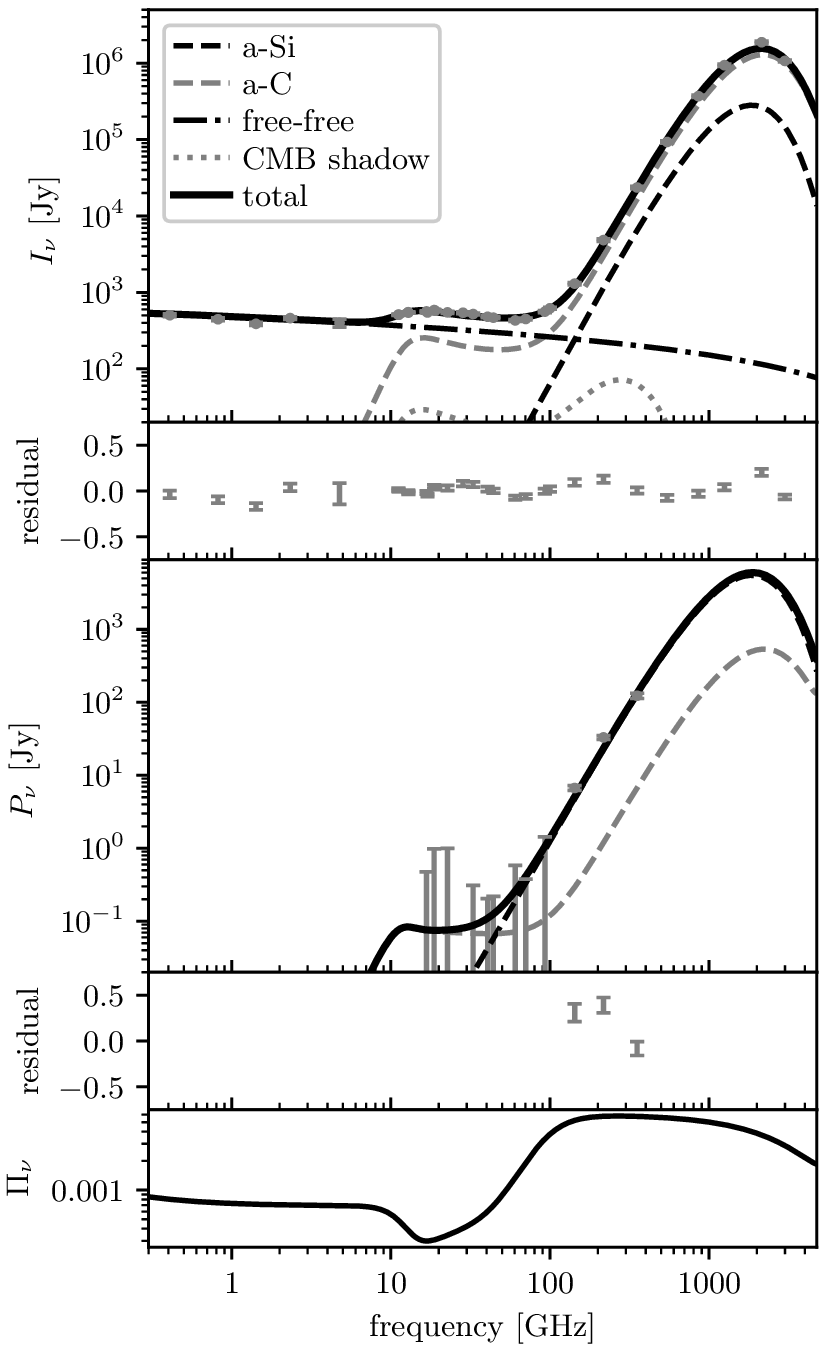}
      \label{fig:SED_W43}
    }
    \caption{
    The intensity and polarization SEDs of Perseus and W\,43 molecular clouds with the best-fit model.
    Absolute values are plotted for the spectra of the CMB temperature anisotropy (black dotted curve) and the CMB shadow (grey dotted curve).
    The lowest panel shows the predicted frequency dependence of the polarization fraction for each molecular cloud.
    }
    \label{fig:SED}
\end{figure*}
\begin{table}
    \centering
    \caption{Best-fit parameters' values}
    \label{tab:best-fit-params}
    \begin{tabular}{lcc}
        \hline\hline
        & Perseus & W\,43 \\ 
        \hline
        $T_\mathrm{a\mathchar`-Si}$ (K) &
        $14.88^{+0.18}_{-0.17}$ & $17.82^{+0.11}_{-0.11}$ \\
        $T_\mathrm{a\mathchar`-C}$ (K) &
        $17.36^{+0.21}_{-0.20}$ & $20.92^{+0.14}_{-0.13}$ \\
        $f_\mathrm{a\mathchar`-Si}^\mathrm{TLS}$ ($\times10^{-4}$) &
        $4.239^{+3.102}_{-2.477}$ & $2.009^{+0.942}_{-0.906}$ \\
        $f_\mathrm{a\mathchar`-C}^\mathrm{TLS}$ ($\times10^{-2}$) &
        $2.174^{+0.692}_{-0.529}$ & $3.754^{+0.293}_{-0.275}$ \\
        $L_\mathrm{a\mathchar`-Si}^\mathrm{min}$ ($\times10^{-1}$) &
        $2.865^{+0.108}_{-0.109}$ & $3.266^{+0.006}_{-0.006}$ \\
        $\delta L_\mathrm{a\mathchar`-C}$ ($\times10^{-5}$) $^\ast$ &
        $97.92^{+102.61}_{-73.89}$ & $6.604^{+6.673}_{-4.918}$ \\
        $\Delta_{0,\mathrm{a\mathchar`-C}}^\mathrm{max}$ (GHz/$h$) &
        $16.03^{+1.69}_{-1.67}$ & $11.21^{+0.26}_{-0.27}$ \\
        $\Delta_{0,\mathrm{a\mathchar`-C}}^\mathrm{min}$ (GHz/$h$) &
        $10.01^{+4.27}_{-3.98}$ & $11.13^{+0.27}_{-0.28}$ \\ 
        $\tau_{+,\mathrm{a\mathchar`-C}}$ ($10^{-11}$\,s) &
        $2.111^{+0.157}_{-0.205}$ & $2.886^{+0.067}_{-0.065}$ \\ 
        $N_\mathrm{d}$ ($10^9$\,cm$^{-2}$) &
        $5.499^{+0.408}_{-0.398}$ & $78.97^{+2.41}_{-2.39}$ \\ 
        EM (pc\,cm$^{-6}$) &
        $26.61^{+4.02}_{-3.97}$ & $4220^{+57}_{-58}$ \\ 
        $\delta T_\mathrm{CMB}$ ($\mu$K) & 
        $-43.54^{+21.60}_{-26.70}$ & ---$^\dagger$
        \\ \hline
    \end{tabular}
    \begin{tablenotes}\footnotesize
        \item[] $\ast$ We define $\delta L_\mathrm{a\mathchar`-C} \equiv 1/3-L_\mathrm{a\mathchar`-C}^\mathrm{min}$.
        \item[] $\dagger$ Since the contribution of the CMB temperature anisotropy is removed from the data of W\,43, $\delta T_\mathrm{CMB}$ is not a fitting parameter.
    \end{tablenotes}
\end{table}

The predicted spectra of thermal emission from the amorphous dust composed of a-Si and a-C are compared with the observed intensity and polarization spectra of Perseus and W\,43 molecular clouds \citep{Nashimoto+2020}.
The contribution of the foreground and background interstellar matter of each molecular cloud has already been removed from these data. 
In addition to the dust emission, the free-free emission attributed to each molecular cloud can be seen in the intensity spectra. 
The frequency dependence of the free-free contributions was modeled by the formula given by \cite{PlanckXX_2011}. 
The emission measures (EMs), which are equivalent to the amplitude of the free-free emission, were treated as fitting free parameters. 
The cosmic microwave background (CMB) temperature anisotropy was already removed from the spectra of the W\,43. 
On the other hand, the CMB temperature anisotropy has not been removed from the spectra of the Perseus.
Therefore, the CMB temperature anisotropy was taken into account when the Perseus intensity spectra were fitted. 
The absorption of the CMB monopole due to interstellar dust, which is named the CMB shadow by \cite{CMBShadow+2020}, is taken into account in the fitting self consistently.
To perform the fitting, we use \textsf{emcee} Markov Chain Monte Carlo software \citep{emcee_2013}.
The means of the probability density distributions for each parameter estimated from the MCMC method are adopted as the values of the best-fit model.

The intensity and polarization spectra of the best-fit model are overlaid on the observed spectra for each molecular cloud in Figure \ref{fig:SED}. 
The best-fit parameters are summarized in Table \ref{tab:best-fit-params}. 
Our model provides an excellent fit to the observed intensity and polarization spectra simultaneously from microwave through far infrared. 
In this model, the AME originates from the resonance emission of the TLS in a-C, and the dominant contributor to the intensity spectra from far infrared to AME is a-C.
The observed polarized emission in the submillimeter waveband is attributed to a-Si. 
Our model predicts that almost all polarized emission is originated from a-Si. 
On the other hand, the shape of a-C is very close to spherical and the polarized radiation emitted from a-C is negligibly small. 
According to \cite{Draine+2009}, the dust model composed of spherical a-C and ellipsoidal a-Si is compatible with all the observables from optical through far infrared. 
Figure \ref{fig:SED} shows the frequency dependence of the expected polarization fraction of dust.
The polarization fraction at 17\,GHz predicted by our model is $5.763\times10^{-3}$ for the Perseus cloud and $2.983\times10^{-4}$ for the W\,43 cloud. 
From the 3$\sigma$ lower limit of $L^\mathrm{min}_\mathrm{a\mathchar`-C}$, the 3$\sigma$ lower limit of the polarization fraction is provided as $8.129\times10^{-5}$ and $8.012\times10^{-6}$, and the ratio of polarized emission of a-C to that of a-Si is 0.1554 and 0.08180 for the Perseus and the W\,43, respectively.
These results are consistent with the observed upper limit (\citealt{QUIJOTE1_2015}; \citeyear{QUIJOTE2_2017}). 
Since the dominant contributor of the polarized emission of the AME is a-C in our model, 
eventually a detection of polarized AME emission would allow to
constrain the degree of the asphericity of a-C that is $L^\mathrm{min}_\mathrm{a\mathchar`-C}$. 
Although \cite{Nashimoto+2020} predicted a 90-degree flip in the polarization direction in AME frequency range for W\,43, it did not happen in the current model.
This originates from the difference of the adopted models to describe the contribution due to lattice vibration. 
\cite{Nashimoto+2020} applied the disordered charge distribution (DCD) model \citep{Schlomann_1964} as a contribution due to lattice vibration, while the model proposed by \cite{Draine+1984} is applied in this study.
The flip of the polarization direction occurs when the sign of the real part of the electric susceptibility of dust are reversed.
The real part of the electric susceptibility due to the resonance transition becomes negative around the resonance frequency. 
When the absolute value of the real part of the electric susceptibility due to the resonance transition is larger than that of other contributions, the real part of the electric susceptibility of the whole dust becomes negative. 
The real part of the electric susceptibility due to the lattice vibration proposed by \cite{Draine+1984} is about one order of magnitude larger than that of the DCD model, and its absolute value is larger than the contribution from the resonance transition in all frequency range. 
Under ideal conditions (uniform magnetic field component perpendicular to the line of sight, perfectly aligned dust grains, no turbulent component of the magnetic fields) our best fitting models show that the shape of a-Si and a-C of Perseus and W\,43 are close to perfect sphere.
Such results are roughly consistent with optical polarimetry on Perseus \citep{Goodman1990} and high resolution polarization map of a dense core in W\,43 at 350\,$\mu$m \citep{Dotson2010}. 
The derived dust column density for Perseus is consistent with the visual extinction extracted from 2MASS \citep{Schnee2008}. 
The visual extinction inferred from the derived dust column density for W\,43 is $A_V=40$ because $A_V = 1.086 \pi a^2 Q_V^\mathrm{ext} N_\mathrm{d}$ where we assume the extinction efficiency at the visual band $Q_V^\mathrm{ext}$ of 1.5. 
This is consistent with $A_V$ extracted from the Planck map \citep{Planck2016}. 
More realistic comparisons with dust grain models taking into account environment effects are discussed in the next section.

\section{Discussion}\label{sec:discuss}
We have shown that the thermal emission of the amorphous dust composed of a-Si and a-C grains provides excellent fit to both the observed intensity and the polarization spectra of molecular clouds simultaneously. 
By taking into account a-C, the model prediction of the polarization fraction of the AME is reduced dramatically compared with the prediction made by \cite{Nashimoto+2020}. 
The AME originates from the resonance transition of the TLS attributed to the a-C with almost spherical shape. 
On the other hand, the observed polarized emission in submillimeter wavebands is coming from a-Si.

The systematic errors brought by adopting relations (\ref{eq:T_a-Si}) and (\ref{eq:T_a-C}) to estimate dust temperatures are discussed. 
Since the emission from a-C is the dominant contributor to the intensity SEDs in our model, the temperature of a-C is defined robustly by the far infrared peak position of the intensity SEDs. 
In Perseus, by adopting the opacity model with the best fit parameters for a-C shown in Table \ref{tab:best-fit-params},
$G_0=2.363$ is obtained to reproduce the best temperature of a-C under the energy balance condition between the radiative heating and cooling.   
This means a reduction of 20\% from the best fit value, $G_0=3.048$, obtained by using Equations (\ref{eq:T_a-Si}) and (\ref{eq:T_a-C}). 
This lower value of $G_0$ translates by an equilibrium temperature of a-Si of 14.24\,K when the opacity model with the best fit parameters for a-Si shown in Table \ref{tab:best-fit-params} are adopted, except the temperature. 
The deduced temperature coincides with the best fit temperature shown in Table \ref{tab:best-fit-params} very well. 
Therefore, we conclude that relations (\ref{eq:T_a-Si}) and (\ref{eq:T_a-C}) can be used as good proxies of dust temperature. 
For W\,43, caution must be paid to use these relations since they apply to optically thin clouds. 
The SED of the ISRF has prominent peak in the near infrared regime \citep{Mathis1983}. 
The near infrared extinction inferred from $A_V=40$ for W\,43 is from a few to 10 magnitudes \citep{Gao2013}. 
To model the thermal state of dust grains in W\,43, one would have to solve for the radiative transfer of the ISRF through the cloud which is out of the framework of the present work. 
However, our treatment is enough to show the potential ability of amorphous dust model to fit intensity and polarization SEDs simultaneously from AME through the far infrared peak.

Since we neglected a reduction of the polarization fraction due to astronomical effects \citep{Hildebrand+1995}, the actual interstellar dust may have a lower value of the shape parameters, $L_i^\mathrm{min}$, than those reported in this paper and may have larger ellipticity.  
Our shape distribution model described in Section \ref{sec:model} includes oblate and prolate spheroids for each value of $L_{x,\mathrm{a\mathchar`-Si}}$.
The axial ratio between major and minor axes of oblate ($a_{x,\mathrm{a\mathchar`-Si}} = a_{y,\mathrm{a\mathchar`-Si}} \ge a_{z,\mathrm{a\mathchar`-Si}}$) and prolate ($a_{x,\mathrm{a\mathchar`-Si}} \ge a_{y,\mathrm{a\mathchar`-Si}} = a_{z,\mathrm{a\mathchar`-Si}}$) spheroids limit when $L_{x,\mathrm{a\mathchar`-Si}} = L_\mathrm{a\mathchar`-Si}^\mathrm{min}$ are shown in Table \ref{tab:axial-ratio}. 
The maximum allowed axial ratios of the oblate (prolate) spheroids in the best-fit models are about 1.4 (1.2) for a-Si and about 1.007 (1.004) for a-C in the Perseus molecular cloud, and about 1.05 (1.03) for a-Si and 1.0005 (1.0003) for a-C in the W\,43 molecular cloud, respectively. 
These results indicate that, in order to make our model compatible with the observed data, the shapes of the interstellar dust are close to spherical. 
However, it is natural to assume that the observed polarization fraction of each molecular cloud suffers a significant reduction due to astronomical attenuation.  
Although the galactic magnetic field is assumed to be perfectly perpendicular to the line of sight in this study, it is certain that there are finite inclination and variation of the magnetic field direction along the line of sight. 
Local turbulence in the magnetic fields may also reduce the polarization fraction. 
Depolarization also occurs due to mixing different components along the line of sight with different polarization directions.
In addition, beam depolarization leads to lower observed polarization fraction, especially in the data provided by QUIJOTE, whose beam widths are 1$^\circ$.
Then, we take the maximum value of the observed polarization fraction of the interstellar dust at 353\,GHz of 0.2 \citep{PlanckXII_2018} as the reference value of the intrinsic polarization fraction of the interstellar dust emission. 
The required shape parameters of a-Si, $L_\mathrm{a\mathchar`-Si}^\mathrm{min}$, and the maximum allowed axial ratios of the oblate and prolate spheroids to realize the polarization fractions of 0.2 at 353\,GHz for each molecular cloud are also summarized in Table \ref{tab:axial-ratio}. 
It shows that the large variety of the shapes are allowed for a-Si if the intrinsic polarization fraction is 0.2.
For comparison, the maximum allowed axial ratios of spheroids when the intrinsic polarization fractions at 353\,GHz are 0.05 and 0.1, are shown in Table \ref{tab:axial-ratio}. 
How the shape constraints on a-C coming from our best-fit models are relaxed by taking into account the astronomical attenuation is estimated as follows. 
The model prediction of the polarization fraction of a-C at 353\,GHz for the Perseus molecular cloud is increased a factor of $0.2/0.046=4.35$, where the possible intrinsic polarization fraction of 0.2 is divided by the observed polarization fraction at 353\,GHz of 0.046. 
The prediction of the polarization fraction of a-C from our best model at 353\,GHz is $6\times10^{-3}$. 
The possible intrinsic polarization fraction of a-C is estimated as 0.026. 
This results in $\delta L_\mathrm{a\mathchar`-C} \equiv 1/3-L_\mathrm{a\mathchar`-C}^\mathrm{min} = 6\times10^{-3}$. 
The maximum allowed axial ratios of a-C are relaxed to 1.04 for oblates and to 1.02 for prolates, respectively. 
Similarly, in the W\,43 molecular cloud, the maximum allowed axial ratios of a-C are relaxed to 1.03 for oblates and to 1.01 for prolates, respectively.
The shape of a-C has to be still close to spherical. 
Imperfectness of the dust grains alignment relative to the magnetic field could further relax the constraint on the dust grain shape. 
It is known that the dust grains are not perfectly aligned relative to the magnetic field (\citealt{Hildebrand+1995}; \citealt{Guillet+2018}).

\cite{Guillet+2018} pointed out although the large grain with $a=0.1\,\mu$m may be aligned almost perfectly, the degree of the alignment decreases as the size of dust decreases.
Although further studies on the astronomical attenuation of the polarization fractions of the dust emission are required to extract the information of dust shape, it is certain that our model predicts that the shape of a-C is close to spherical and a-Si has variety of ellipticity.
\begin{table}
    \centering
    \caption{Maximum axial ratio of a-Si}
    \label{tab:axial-ratio}
    \begin{tabular}{lcccccc}
        \hline\hline
        & \multicolumn{3}{c}{Perseus} & \multicolumn{3}{c}{W\,43} \\ 
        & $L^\mathrm{min}_\mathrm{a\mathchar`-Si}$ & Oblate & Prolate 
        & $L^\mathrm{min}_\mathrm{a\mathchar`-Si}$ & Oblate & Prolate \\
        \hline 
        best-fit model & 0.2865 & 1.40 & 1.20 & 0.3266 & 1.05 & 1.03 \\
        $\Pi_\mathrm{353GHz}=0.05$ & 0.273 & 1.54 & 1.26 & 0.268 & 1.60 & 1.29 \\
        $\Pi_\mathrm{353GHz}=0.1$ & 0.207 & 2.48 & 1.69 & 0.202 & 2.57 & 1.73 \\
        $\Pi_\mathrm{353GHz}=0.2$ & 0.102 & 6.41 & 3.16 & 0.0996 & 6.59 & 3.22
        \\ \hline
    \end{tabular}
\end{table}

\begin{figure}[t!]
    \centering
    \includegraphics{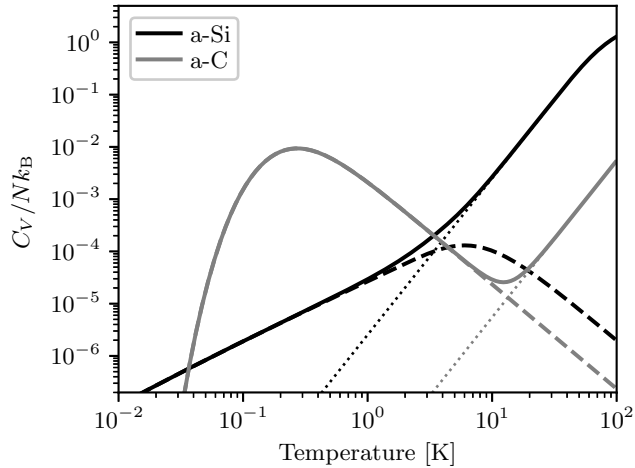}
    \caption{
    The predicted heat capacity of each amorphous dust species in Perseus molecular cloud.
    The dashed curves and the dotted curves are contribution from the TLS model and the Debye model, respectively, and solid curves are sum of them.
    }
    \label{fig:Cv}
\end{figure}

Figure \ref{fig:Cv} shows the temperature dependence of the heat capacity of a-C and a-Si with the best-fit parameters for the Perseus cloud summarized in Table \ref{tab:best-fit-params}. 
Below a few kelvins, the contribution from the TLS becomes dominant. 
The heat capacity of a-Si shows a linear dependence on temperature as observed in the laboratory experiments.
The heat capacity of a-C shows a bump at around sub kelvin. 
The heat capacity of a single TLS is described by a function called a Schottky heat capacity, which has a peak at $k_\mathrm{B}T_i \simeq 0.42 E$.
The temperature dependence of the heat capacity of the amorphous material is defined by the superposition of each TLS in the material, which has different $E$. 
The distribution of tunnel splitting energy $\Delta_0$ of a-C is limited to a narrow range in order to reproduce the intensity of AME.
As a result, the distribution of $E$ is also limited in a narrow range.
This is the reason why the temperature dependence of a-C shows such peculiar characteristics. 
Our model prediction of the fraction of the atoms trapped in the TLS, $f^\mathrm{TLS}_\mathrm{a\mathchar`-Si}$, of a-Si is the order of comparable to the laboratory measurements for the amorphous material \citep{Phillips_1987}.
On the other hand, our model prediction of $f^\mathrm{TLS}_\mathrm{a\mathchar`-C}$ is about two orders of magnitude larger than that of a-Si.
Because of this, the heat capacity of a-C predicted by the TLS model is much larger than the Debye heat capacity around 0.1\,K. 
Since little has been done for the laboratory measurement of the heat capacity of the carbonaceous amorphous material, measuring the low-temperature behavior of the heat capacity of a-C in the laboratory is important to test our prediction. 
Another possibility is that the characteristics of the amorphous dust predicted by our model are specific to the interstellar dust grains.  
It does not affect the far infrared emission and AME since 
such a low temperature behavior of the heat capacity has little influence on the thermal history of big grains. 
Supposing the latter possibility, we name our model as the cosmic amorphous dust model, abbreviated to CAD. 
Testing CAD possibility of the origin of AME by laboratory experiments and astronomical observations is worth to do.

\acknowledgments
We would like to 
thank Satoshi Tomita for fruitful discussions and anonymous 
referee for ones constructive comments.
This work is supported by MEXT KAKENHI Grant Number 18H05539.
This work is partially supported by the Graduate Program on Physics for the Universe (GP-PU), Tohoku University.
FP acknowledges support from the Spanish Ministerio de Ciencia, Innovación y Universidades (MICINN) under grant numbers ESP2015-65597-C4-4-R and ESP2017-86852-C4-2-R.


\typeout{\the\interdisplaylinepenalty}
\bibliography{article}{}

\begin{thebibliography}{}
\expandafter\ifx\csname natexlab\endcsname\relax\def\natexlab#1{#1}\fi
\providecommand{\url}[1]{\href{#1}{#1}}
\providecommand{\dodoi}[1]{doi:~\href{http://doi.org/#1}{\nolinkurl{#1}}}
\providecommand{\doeprint}[1]{\href{http://ascl.net/#1}{\nolinkurl{http://ascl.net/#1}}}
\providecommand{\doarXiv}[1]{\href{https://arxiv.org/abs/#1}{\nolinkurl{https://arxiv.org/abs/#1}}}

\bibitem[{{Anderson} {et~al.}(1972){Anderson}, {Halperin}, \&
  {Varma}}]{Anderson+1972}
{Anderson}, P.~W., {Halperin}, B.~I., \& {Varma}, C.~M. 1972, Philosophical
  Magazine, 25, 1, \dodoi{10.1080/14786437208229210}

\bibitem[{{Bohren} \& {Huffman}(1983)}]{Bohren+1983}
{Bohren}, C.~F., \& {Huffman}, D.~R. 1983, {Absorption and scattering of light
  by small particles} (New York: Wiley).
\newblock \url{https://books.google.co.jp/books?id=S1RCZ8BjgN0C}

\bibitem[{{B{\"o}sch}(1978)}]{Bosch_1978}
{B{\"o}sch}, M.~A. 1978, \prl, 40, 879, \dodoi{10.1103/PhysRevLett.40.879}

\bibitem[{{Davies} {et~al.}(2006){Davies}, {Dickinson}, {Banday}, {Jaffe},
  {G{\'o}rski}, \& {Davis}}]{Davis+2006}
{Davies}, R.~D., {Dickinson}, C., {Banday}, A.~J., {et~al.} 2006, \mnras, 370,
  1125, \dodoi{10.1111/j.1365-2966.2006.10572.x}

\bibitem[{{Dickinson} {et~al.}(2018){Dickinson}, {Ali-Ha{\"\i}moud}, {Barr},
  {Battistelli}, {Bell}, {Bernstein}, {Casassus}, {Cleary}, {Draine},
  {G{\'e}nova-Santos}, {Harper}, {Hensley}, {Hill-Valler}, {Hoang}, {Israel},
  {Jew}, {Lazarian}, {Leahy}, {Leech}, {L{\'o}pez-Caraballo}, {McDonald},
  {Murphy}, {Onaka}, {Paladini}, {Peel}, {Perrott}, {Poidevin}, {Readhead},
  {Rubi{\~n}o-Mart{\'\i}n}, {Taylor}, {Tibbs}, {Todorovi{\'c}}, \&
  {Vidal}}]{Dickinson+2018}
{Dickinson}, C., {Ali-Ha{\"\i}moud}, Y., {Barr}, A., {et~al.} 2018, New
  Astronomy Reviews, 80, 1, \dodoi{10.1016/j.newar.2018.02.001}

\bibitem[{{Dotson} {et~al.}(2010){Dotson}, {Vaillancourt}, {Kirby}, {Dowell},
  {Hildebra}, \& {Davidson}}]{Dotson2010}
{Dotson}, J., {Vaillancourt}, J., {Kirby}, L., {et~al.} 2010, \apjs, 186, 406,
  \dodoi{10.1088/0067-0049/186/2/406}

\bibitem[{{Draine} \& {Fraisse}(2009)}]{Draine+2009}
{Draine}, B.~T., \& {Fraisse}, A.~A. 2009, \apj, 696, 1,
  \dodoi{10.1088/0004-637X/696/1/1}

\bibitem[{{Draine} \& {Hensley}(2017)}]{Draine+2017}
{Draine}, B.~T., \& {Hensley}, B.~S. 2017, arXiv:1710.08968.
\newblock \doarXiv{1710.08968}

\bibitem[{{Draine} \& {Lee}(1984)}]{Draine+1984}
{Draine}, B.~T., \& {Lee}, H.~M. 1984, \apj, 285, 89, \dodoi{10.1086/162480}

\bibitem[{{Foreman-Mackey} {et~al.}(2013){Foreman-Mackey}, {Hogg}, {Lang}, \&
  {Goodman}}]{emcee_2013}
{Foreman-Mackey}, D., {Hogg}, D.~W., {Lang}, D., \& {Goodman}, J. 2013, \pasp,
  125, 306, \dodoi{10.1086/670067}

\bibitem[{{Gao} {et~al.}(2013){Gao}, {Li}, \& {Jiang}}]{Gao2013}
{Gao}, J., {Li}, A., \& {Jiang}, B. 2013, Earth, Plants and Space, 65, 1127,
  \dodoi{10.5047/eps.2013.05.016}

\bibitem[{{G{\'e}nova-Santos} {et~al.}(2015){G{\'e}nova-Santos},
  {Rubi{\~n}o-Mart{\'\i}n}, {Rebolo}, {Pel{\'a}ez-Santos},
  {L{\'o}pez-Caraballo}, {Harper}, {Watson}, {Ashdown}, {Barreiro},
  {Casaponsa}, {Dickinson}, {Diego}, {Fern{\'a}ndez-Cobos}, {Grainge},
  {Guti{\'e}rrez}, {Herranz}, {Hoyland}, {Lasenby}, {L{\'o}pez-Caniego},
  {Mart{\'\i}nez-Gonz{\'a}lez}, {McCulloch}, {Melhuish}, {Piccirillo},
  {Perrott}, {Poidevin}, {Razavi-Ghods}, {Scott}, {Titterington}, {Tramonte},
  {Vielva}, \& {Vignaga}}]{QUIJOTE1_2015}
{G{\'e}nova-Santos}, R., {Rubi{\~n}o-Mart{\'\i}n}, J.~A., {Rebolo}, R.,
  {et~al.} 2015, \mnras, 452, 4169, \dodoi{10.1093/mnras/stv1405}

\bibitem[{{G{\'e}nova-Santos} {et~al.}(2017){G{\'e}nova-Santos},
  {Rubi{\~n}o-Mart{\'\i}n}, {Pel{\'a}ez-Santos}, {Poidevin}, {Rebolo},
  {Vignaga}, {Artal}, {Harper}, {Hoyland}, {Lasenby},
  {Mart{\'\i}nez-Gonz{\'a}lez}, {Piccirillo}, {Tramonte}, \&
  {Watson}}]{QUIJOTE2_2017}
{G{\'e}nova-Santos}, R., {Rubi{\~n}o-Mart{\'\i}n}, J.~A., {Pel{\'a}ez-Santos},
  A., {et~al.} 2017, \mnras, 464, 4107, \dodoi{10.1093/mnras/stw2503}

\bibitem[{{Goodman} {et~al.}(1990){Goodman}, {Bastien}, {Myers}, \&
  {M\'enard}}]{Goodman1990}
{Goodman}, A., {Bastien}, P., {Myers}, P., \& {M\'enard}, F. 1990, \apj, 359,
  363, \dodoi{10.1086/169070}

\bibitem[{{Guillet} {et~al.}(2018){Guillet}, {Fanciullo}, {Verstraete},
  {Boulanger}, {Jones}, {Miville-Desch{\^e}nes}, {Ysard}, {Levrier}, \&
  {Alves}}]{Guillet+2018}
{Guillet}, V., {Fanciullo}, L., {Verstraete}, L., {et~al.} 2018, \aap, 610,
  A16, \dodoi{10.1051/0004-6361/201630271}

\bibitem[{{Hildebrand} \& {Dragovan}(1995)}]{Hildebrand+1995}
{Hildebrand}, R.~H., \& {Dragovan}, M. 1995, \apj, 450, 663,
  \dodoi{10.1086/176173}

\bibitem[{{Hirashita} \& {Yan}(2009)}]{Hirashita+2009}
{Hirashita}, H., \& {Yan}, H. 2009, \mnras, 394, 1061,
  \dodoi{10.1111/j.1365-2966.2009.14405.x}

\bibitem[{{Jones}(2009)}]{Jones_2009}
{Jones}, A.~P. 2009, \aap, 506, 797, \dodoi{10.1051/0004-6361/200810621}

\bibitem[{{Li} \& {Draine}(2001{\natexlab{a}})}]{LiDraine2001}
{Li}, A., \& {Draine}, B.~T. 2001{\natexlab{a}}, \apjl, 550, L213,
  \dodoi{10.1086/319640}

\bibitem[{{Li} \& {Draine}(2001{\natexlab{b}})}]{Li+2001}
---. 2001{\natexlab{b}}, \apj, 554, 778, \dodoi{10.1086/323147}

\bibitem[{{Mathis} {et~al.}(1983){Mathis}, {Mezger}, \& {Panagia}}]{Mathis1983}
{Mathis}, J.~S., {Mezger}, P.~G., \& {Panagia}, N. 1983, \aa, 128, 212

\bibitem[{{Meny} {et~al.}(2007){Meny}, {Gromov}, {Boudet}, {Bernard},
  {Paradis}, \& {Nayral}}]{Meny+2007}
{Meny}, C., {Gromov}, V., {Boudet}, N., {et~al.} 2007, \aap, 468, 171,
  \dodoi{10.1051/0004-6361:20065771}

\bibitem[{{Mishra} \& {Li}(2015)}]{MishraLi2015}
{Mishra}, A., \& {Li}, A. 2015, \apj, 809, 120(13pp),
  \dodoi{10.1088/0004-637X/809/2/120}

\bibitem[{{Nashimoto} {et~al.}(2020{\natexlab{a}}){Nashimoto}, {Hattori}, \&
  {Chinone}}]{CMBShadow+2020}
{Nashimoto}, M., {Hattori}, M., \& {Chinone}, Y. 2020{\natexlab{a}}, \apjl,
  895, L21, \dodoi{10.3847/2041-8213/ab9018}

\bibitem[{{Nashimoto} {et~al.}(2020{\natexlab{b}}){Nashimoto}, {Hattori},
  {G{\'e}nova-Santos}, \& {Poidevin}}]{Nashimoto+2020}
{Nashimoto}, M., {Hattori}, M., {G{\'e}nova-Santos}, R., \& {Poidevin}, F.
  2020{\natexlab{b}}, \pasj, 72, 6, \dodoi{10.1093/pasj/psz124}

\bibitem[{Nittke {et~al.}(1998)Nittke, Sahling, \& Esquinazi}]{Nittke+1998}
Nittke, A., Sahling, S., \& Esquinazi, P. 1998, Heat Release in Solids, ed.
  P.~Esquinazi (Berlin, Heidelberg: Springer Berlin Heidelberg), 9--56,
  \dodoi{10.1007/978-3-662-03695-2_2}

\bibitem[{{Paradis} {et~al.}(2011){Paradis}, {Bernard}, {M{\'e}ny}, \&
  {Gromov}}]{Paradis+2011}
{Paradis}, D., {Bernard}, J.-P., {M{\'e}ny}, C., \& {Gromov}, V. 2011, \aap,
  534, A118, \dodoi{10.1051/0004-6361/201116862}

\bibitem[{{Phillips}(1972)}]{Phillips_1972}
{Phillips}, W.~A. 1972, Journal of Low Temperature Physics, 7, 351,
  \dodoi{10.1007/BF00660072}

\bibitem[{{Phillips}(1987)}]{Phillips_1987}
---. 1987, Reports on Progress in Physics, 50, 1657,
  \dodoi{10.1088/0034-4885/50/12/003}

\bibitem[{{Planck Collaboration} {et~al.}(2011){Planck Collaboration}, {Ade},
  {Aghanim}, {Arnaud}, {Ashdown}, {Aumont}, {Baccigalupi}, {Balbi}, {Banday},
  {Barreiro}, {Bartlett}, {Battaner}, {Benabed}, {Beno{\^\i}t}, {Bernard},
  {Bersanelli}, {Bhatia}, {Bock}, {Bonaldi}, {Bond}, {Borrill}, {Bouchet},
  {Boulanger}, {Bucher}, {Burigana}, {Cabella}, {Cappellini}, {Cardoso},
  {Casassus}, {Catalano}, {Cay{\'o}n}, {Challinor}, {Chamballu}, {Chary},
  {Chen}, {Chiang}, {Chiang}, {Christensen}, {Clements}, {Colombi}, {Couchot},
  {Coulais}, {Crill}, {Cuttaia}, {Danese}, {Davies}, {Davis}, {de Bernardis},
  {de Gasperis}, {de Rosa}, {de Zotti}, {Delabrouille}, {Delouis}, {Dickinson},
  {Donzelli}, {Dor{\'e}}, {D{\"o}rl}, {Douspis}, {Dupac}, {Efstathiou},
  {En{\ss}lin}, {Eriksen}, {Finelli}, {Forni}, {Frailis}, {Franceschi},
  {Galeotta}, {Ganga}, {G{\'e}nova-Santos}, {Giard}, {Giardino},
  {Giraud-H{\'e}raud}, {Gonz{\'a}lez-Nuevo}, {G{\'o}rski}, {Gratton},
  {Gregorio}, {Gruppuso}, {Hansen}, {Harrison}, {Helou}, {Henrot-Versill{\'e}},
  {Herranz}, {Hildebrandt}, {Hivon}, {Hobson}, {Holmes}, {Hovest}, {Hoyland},
  {Huffenberger}, {Jaffe}, {Jaffe}, {Jones}, {Juvela}, {Keih{\"a}nen},
  {Keskitalo}, {Kisner}, {Kneissl}, {Knox}, {Kurki-Suonio}, {Lagache},
  {L{\"a}hteenm{\"a}ki}, {Lamarre}, {Lasenby}, {Laureijs}, {Lawrence}, {Leach},
  {Leonardi}, {Lilje}, {Linden-V{\o}rnle}, {L{\'o}pez-Caniego}, {Lubin},
  {Mac{\'\i}as-P{\'e}rez}, {MacTavish}, {Maffei}, {Maino}, {Mandolesi}, {Mann},
  {Maris}, {Marshall}, {Mart{\'\i}nez-Gonz{\'a}lez}, {Masi}, {Matarrese},
  {Matthai}, {Mazzotta}, {McGehee}, {Meinhold}, {Melchiorri}, {Mendes},
  {Mennella}, {Mitra}, {Miville-Desch{\^e}nes}, {Moneti}, {Montier},
  {Morgante}, {Mortlock}, {Munshi}, {Murphy}, {Naselsky}, {Natoli},
  {Netterfield}, {N{\o}rgaard-Nielsen}, {Noviello}, {Novikov}, {Novikov},
  {O'Dwyer}, {Osborne}, {Pajot}, {Paladini}, {Partridge}, {Pasian},
  {Patanchon}, {Pearson}, {Peel}, {Perdereau}, {Perotto}, {Perrotta},
  {Piacentini}, {Piat}, {Plaszczynski}, {Platania}, {Pointecouteau}, {Polenta},
  {Ponthieu}, {Poutanen}, {Pr{\'e}zeau}, {Procopio}, {Prunet}, {Puget},
  {Reach}, {Rebolo}, {Reich}, {Reinecke}, {Renault}, {Ricciardi}, {Riller},
  {Ristorcelli}, {Rocha}, {Rosset}, {Rowan-Robinson}, {Rubi{\~n}o-Mart{\'\i}n},
  {Rusholme}, {Sandri}, {Santos}, {Savini}, {Scott}, {Seiffert}, {Shellard},
  {Smoot}, {Starck}, {Stivoli}, {Stolyarov}, {Stompor}, {Sudiwala}, {Sygnet},
  {Tauber}, {Terenzi}, {Toffolatti}, {Tomasi}, {Torre}, {Tristram}, {Tuovinen},
  {Umana}, {Valenziano}, {Varis}, {Verstraete}, {Vielva}, {Villa}, {Vittorio},
  {Wade}, {Wandelt}, {Watson}, {Wilkinson}, {Ysard}, {Yvon}, {Zacchei}, \&
  {Zonca}}]{PlanckXX_2011}
{Planck Collaboration}, {Ade}, P.~A.~R., {Aghanim}, N., {et~al.} 2011, \aap,
  536, A20, \dodoi{10.1051/0004-6361/201116470}

\bibitem[{{Planck Collaboration} {et~al.}(2016){Planck Collaboration}, {Ade},
  {Aghanim}, {Arnaud}, {Ashdown}, {Aumont}, {Baccigalupi}, {Balbi}, {Banday},
  {Barreiro}, {Bartlett}, {Battaner}, {Benabed}, {Beno{\^\i}t}, {Bernard},
  {Bersanelli}, {Bhatia}, {Bock}, {Bonaldi}, {Bond}, {Borrill}, {Bouchet},
  {Boulanger}, {Bucher}, {Burigana}, {Cabella}, {Cappellini}, {Cardoso},
  {Casassus}, {Catalano}, {Cay{\'o}n}, {Challinor}, {Chamballu}, {Chary},
  {Chen}, {Chiang}, {Chiang}, {Christensen}, {Clements}, {Colombi}, {Couchot},
  {Coulais}, {Crill}, {Cuttaia}, {Danese}, {Davies}, {Davis}, {de Bernardis},
  {de Gasperis}, {de Rosa}, {de Zotti}, {Delabrouille}, {Delouis}, {Dickinson},
  {Donzelli}, {Dor{\'e}}, {D{\"o}rl}, {Douspis}, {Dupac}, {Efstathiou},
  {En{\ss}lin}, {Eriksen}, {Finelli}, {Forni}, {Frailis}, {Franceschi},
  {Galeotta}, {Ganga}, {G{\'e}nova-Santos}, {Giard}, {Giardino},
  {Giraud-H{\'e}raud}, {Gonz{\'a}lez-Nuevo}, {G{\'o}rski}, {Gratton},
  {Gregorio}, {Gruppuso}, {Hansen}, {Harrison}, {Helou}, {Henrot-Versill{\'e}},
  {Herranz}, {Hildebrandt}, {Hivon}, {Hobson}, {Holmes}, {Hovest}, {Hoyland},
  {Huffenberger}, {Jaffe}, {Jaffe}, {Jones}, {Juvela}, {Keih{\"a}nen},
  {Keskitalo}, {Kisner}, {Kneissl}, {Knox}, {Kurki-Suonio}, {Lagache},
  {L{\"a}hteenm{\"a}ki}, {Lamarre}, {Lasenby}, {Laureijs}, {Lawrence}, {Leach},
  {Leonardi}, {Lilje}, {Linden-V{\o}rnle}, {L{\'o}pez-Caniego}, {Lubin},
  {Mac{\'\i}as-P{\'e}rez}, {MacTavish}, {Maffei}, {Maino}, {Mandolesi}, {Mann},
  {Maris}, {Marshall}, {Mart{\'\i}nez-Gonz{\'a}lez}, {Masi}, {Matarrese},
  {Matthai}, {Mazzotta}, {McGehee}, {Meinhold}, {Melchiorri}, {Mendes},
  {Mennella}, {Mitra}, {Miville-Desch{\^e}nes}, {Moneti}, {Montier},
  {Morgante}, {Mortlock}, {Munshi}, {Murphy}, {Naselsky}, {Natoli},
  {Netterfield}, {N{\o}rgaard-Nielsen}, {Noviello}, {Novikov}, {Novikov},
  {O'Dwyer}, {Osborne}, {Pajot}, {Paladini}, {Partridge}, {Pasian},
  {Patanchon}, {Pearson}, {Peel}, {Perdereau}, {Perotto}, {Perrotta},
  {Piacentini}, {Piat}, {Plaszczynski}, {Platania}, {Pointecouteau}, {Polenta},
  {Ponthieu}, {Poutanen}, {Pr{\'e}zeau}, {Procopio}, {Prunet}, {Puget},
  {Reach}, {Rebolo}, {Reich}, {Reinecke}, {Renault}, {Ricciardi}, {Riller},
  {Ristorcelli}, {Rocha}, {Rosset}, {Rowan-Robinson}, {Rubi{\~n}o-Mart{\'\i}n},
  {Rusholme}, {Sandri}, {Santos}, {Savini}, {Scott}, {Seiffert}, {Shellard},
  {Smoot}, {Starck}, {Stivoli}, {Stolyarov}, {Stompor}, {Sudiwala}, {Sygnet},
  {Tauber}, {Terenzi}, {Toffolatti}, {Tomasi}, {Torre}, {Tristram}, {Tuovinen},
  {Umana}, {Valenziano}, {Varis}, {Verstraete}, {Vielva}, {Villa}, {Vittorio},
  {Wade}, {Wandelt}, {Watson}, {Wilkinson}, {Ysard}, {Yvon}, {Zacchei}, \&
  {Zonca}}]{Planck2016}
---. 2016, \aa, 586, A132, \dodoi{10.1051/0004-6361/201424945}

\bibitem[{{Planck Collaboration} {et~al.}(2018){Planck Collaboration},
  {Aghanim}, {Akrami}, {Alves}, {Ashdown}, {Aumont}, {Baccigalupi},
  {Ballardini}, {Banday}, {Barreiro}, {Bartolo}, {Basak}, {Benabed}, {Bernard},
  {Bersanelli}, {Bielewicz}, {Bock}, {Bond}, {Borrill}, {Bouchet}, {Boulanger},
  {Bracco}, {Bucher}, {Burigana}, {Calabrese}, {Cardoso}, {Carron}, {Chary},
  {Chiang}, {Colombo}, {Combet}, {Crill}, {Cuttaia}, {de Bernardis}, {de
  Zotti}, {Delabrouille}, {Delouis}, {Di Valentino}, {Dickinson}, {Diego},
  {Dor{\'e}}, {Douspis}, {Ducout}, {Dupac}, {Efstathiou}, {Elsner},
  {En{\ss}lin}, {Eriksen}, {Falgarone}, {Fantaye}, {Fernandez-Cobos},
  {Ferri{\`e}re}, {Finelli}, {Forastieri}, {Frailis}, {Fraisse}, {Franceschi},
  {Frolov}, {Galeotta}, {Galli}, {Ganga}, {G{\'e}nova-Santos}, {Gerbino},
  {Ghosh}, {Gonz{\'a}lez-Nuevo}, {G{\'o}rski}, {Gratton}, {Green}, {Gruppuso},
  {Gudmundsson}, {Guillet}, {Handley}, {Hansen}, {Helou}, {Herranz}, {Hivon},
  {Huang}, {Jaffe}, {Jones}, {Keih{\"a}nen}, {Keskitalo}, {Kiiveri}, {Kim},
  {Krachmalnicoff}, {Kunz}, {Kurki-Suonio}, {Lagache}, {Lamarre}, {Lasenby},
  {Lattanzi}, {Lawrence}, {Le Jeune}, {Levrier}, {Liguori}, {Lilje},
  {Lindholm}, {L{\'o}pez-Caniego}, {Lubin}, {Ma}, {Mac{\'\i}as-P{\'e}rez},
  {Maggio}, {Maino}, {Mandolesi}, {Mangilli}, {Marcos-Caballero}, {Maris},
  {Martin}, {Mart{\'\i}nez-Gonz{\'a}lez}, {Matarrese}, {Mauri}, {McEwen},
  {Melchiorri}, {Mennella}, {Migliaccio}, {Miville-Desch{\^e}nes}, {Molinari},
  {Moneti}, {Montier}, {Morgante}, {Moss}, {Natoli}, {Pagano}, {Paoletti},
  {Patanchon}, {Perrotta}, {Pettorino}, {Piacentini}, {Polastri}, {Polenta},
  {Puget}, {Rachen}, {Reinecke}, {Remazeilles}, {Renzi}, {Ristorcelli},
  {Rocha}, {Rosset}, {Roudier}, {Rubi{\~n}o-Mart{\'\i}n}, {Ruiz-Granados},
  {Salvati}, {Sandri}, {Savelainen}, {Scott}, {Sirignano}, {Sunyaev},
  {Suur-Uski}, {Tauber}, {Tavagnacco}, {Tenti}, {Toffolatti}, {Tomasi},
  {Trombetti}, {Valiviita}, {Vansyngel}, {Van Tent}, {Vielva}, {Villa},
  {Vittorio}, {Wandelt}, {Wehus}, {Zacchei}, \& {Zonca}}]{PlanckXII_2018}
{Planck Collaboration}, {Aghanim}, N., {Akrami}, Y., {et~al.} 2018, arXiv
  e-prints, arXiv:1807.06212.
\newblock \doarXiv{1807.06212}

\bibitem[{{Schl{\"o}mann}(1964)}]{Schlomann_1964}
{Schl{\"o}mann}, E. 1964, Physical Review, 135, 413,
  \dodoi{10.1103/PhysRev.135.A413}

\bibitem[{{Schnee} {et~al.}(2008){Schnee}, {Li}, {Goodman}, \&
  {Sargent}}]{Schnee2008}
{Schnee}, S., {Li}, J., {Goodman}, A., \& {Sargent}, A. 2008, \apj, 684, 1228,
  \dodoi{10.1086/421339}

\bibitem[{{Tielens}(2005)}]{Tielens_2005}
{Tielens}, A.~G.~G.~M. 2005, {The Physics and Chemistry of the Interstellar
  Medium} (Cambridge University Press)

\bibitem[{{Wei} {et~al.}(2005){Wei}, {Wang}, \& {Wang}}]{Wei+2005}
{Wei}, Y.~X., {Wang}, R.~J., \& {Wang}, W.~H. 2005, \prb, 72, 012203,
  \dodoi{10.1103/PhysRevB.72.012203}

\bibitem[{{Weingartner} \& {Draine}(2001)}]{Weingartner+2001}
{Weingartner}, J.~C., \& {Draine}, B.~T. 2001, \apj, 548, 296,
  \dodoi{10.1086/318651}

\bibitem[{{Zeller} \& {Pohl}(1971)}]{Zeller+1971}
{Zeller}, R.~C., \& {Pohl}, R.~O. 1971, Physical Review B, 4, 2029,
  \dodoi{10.1103/PhysRevB.4.2029}

\end{thebibliography}
\bibliographystyle{aasjournal}
\end{document}